\documentstyle[12pt,epsf]{article}
\newcommand{\be}{\begin{equation}}
\newcommand{\ee}{\end{equation}}
\newcommand{\ba}{\begin{array}}
\newcommand{\ea}{\end{array}}
\newcommand{\bea}{\begin{eqnarray}}
\newcommand{\eea}{\end{eqnarray}}

\begin{document}

\title{\bf Method for control gas diffusion and bubbles formation in liquid porosimetry}
\author{{\bf Dr. V. Smirichinski}\footnote{E-mail:
smirvi@yahoo.com}}
\date{}
 \maketitle

\begin{abstract}
The main problem in liquid porosimetry, which prevents to see the
pore sizes smaller than $2$ microns in diameter, is direct gas
diffusion flow through a micro-porous membrane. This diffusion
causes bubbles formation below the membrane and that spoils
extrusion  (intrusion) data, as one cannot distinguish the volume
of extrusion (intrusion) liquid from the volume of formatted
bubbles. The suggested below method cures the problem by creating
the liquid flow below the membrane. The flow washes out all of the
small bubbles preventing them to grow. That allows using the
membrane at higher differential pressures, even higher than
minimum bubble point of the membrane, without spoiling data.
\end{abstract}

\newpage

\section{Description}
\subsection* {A. Background of the invention}
\noindent

{\bf{A1. Field of the invention}}\\[1mm]

The present invention relates to the field of
porosimetry, particularly to the liquid (not mercury) extrusion or
intrusion porosimeter. In suggested method the porosity is
measured by detecting volume of liquid extruded from (or intruded
in) the sample by applying step by step increasing (decreasing)
gas pressure. The sample is placed on a porous membrane with
pore-size less than the pore-size of the sample.\\[1mm]

{\bf{A2. Description of the previous art}}\\[1mm]

The liquid porosimeter, described by Bernard Miller and Ilya
Tyomkin \cite{MT}, is shown schematically in the Fig. 1.

\begin{figure}[htb]
\vspace*{0cm} \hspace*{1cm} \centerline{ \vbox{ \epsfysize=50mm
\epsfbox{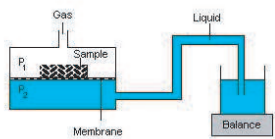} } } \vspace*{-0.3cm}\caption{Basic arrangement
for liquid porosmetry.} \label{fig:1}
\end{figure}

It can be used in both liquid extrusion porosimetry (LEP) and
liquid intrusion porosimetry (LIP). For the LEP mode a
presaturated sample is placed on a micro-porous membrane (membrane
is supported by a rigid porous plate). The gas pressure is
increased in steps and that causes liquid to extrude from the
pores. The largest pores extrude first. The top-loading balance
for each pressure step measures the liquid out-coming from the
sample. The final data for analyzes could be represented as a
function relation $V(DP)$, i.e. extruded volume versus
differential pressure ($DP= P1 - P2)$. Assuming that all of the
through-pores are cylindrical we can apply Laplas equation $d=4 g
\cos(q)/ DP$ to convert the final data to the form $V(d)$.

The same principal works for LIP mode, with the difference in
starting at high pressure and decreasing it stepwise; and, of
course, the sample is not initially saturated. Also LIP test can
be run just after LEP test to see for example liquid
extrusion-intrusion hysteresis. Different liquids can be used in
LEP and LIP tests. The only requirements are the following: liquid
must wet the sample and membrane; the contact angle for the system
of sample-liquid-gas must be known; liquid should be stable and
should have relatively low viscosity. That is the principle of a
liquid porosimetry test.

One of the technical problems in LEP and LIP is the control of
pressure steps during a test. B. Miller and I. Tyomkin  \cite{MT}
have proposed several innovations to control pressure changes more
accurately. Standard scheme of liquid porosimetry allows testing
pore sizes from $2000$ to $2$ micron in diameter. To see smaller
pores the differential pressure should be increased and as it was
mentioned in \cite{RM} to see the smaller pores one need to
increase the pressure hyperbolically. Also the smaller pores have
smaller flow rate through and that requires a longer exposure-
time for high pressure. Under these specific conditions the gas
diffusion flow causes the major problem -- the bubble formations
below the membrane. The volume of the bubbles cannot be
distinguished from the volume of the liquid extruded from
(intruded in) the sample and that spoils the final data when one
tries to see the smallest pores. At this point I will discuss the
problem in more details.

\subsection*{B. Problem with standard liquid porosimetry}

To see the mentioned above gas diffusion clearly, it is enough to
run LEP test without a sample (just a membrane alone). If you
neglect the effect of diffusion (hypothetical situation) the final
data must look like data 1(ideal graph) on Fig. 2.

\begin{figure}[htb]
\vspace*{-0.5cm} \hspace*{-0.5cm} \centerline{ \vbox{
\epsfysize=75mm \epsfbox{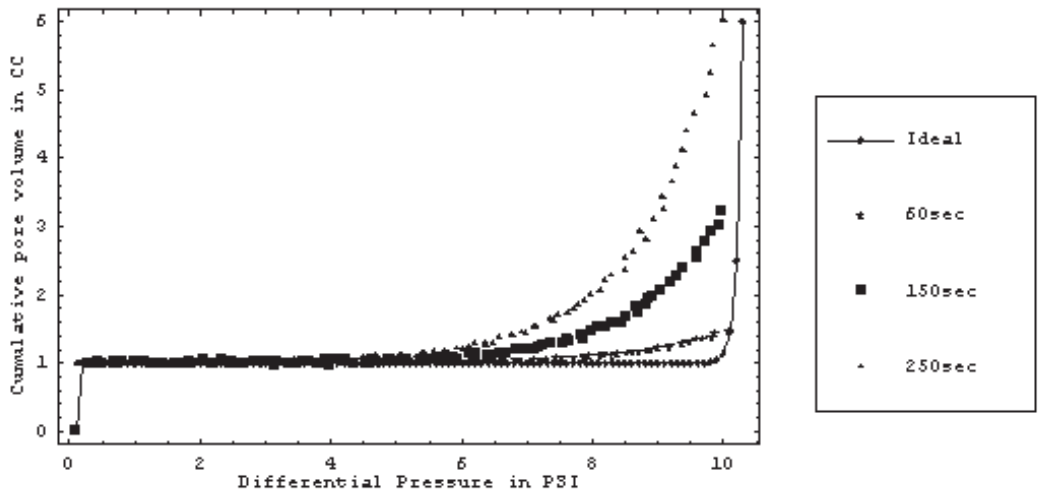} } }
\vspace*{-0.3cm}\caption{Data on membrane test in LEP with
different time-steps (here the time in second denotes the total
time spent for each pressure step). The bubble point for the
membrane is $10.1$ PSI. } \label{fig:2}
\end{figure}

But in real test the data looks like graphs 2-4 on Fig 2. One can
see that the longest in time exposure under pressure causes the
largest deflection from theoretical curve (ideal graph) and that
is due the bubbles formatted below the membrane. To prove the last
statement it is enough to perform some changes in LEP scheme.

Let's do the test with little modification as it shown
schematically on Fig. 3.

The difference with standard liquid porosimeter is that the pump
creates a flow below the membrane. That flow washes out the
bubbles. And if all of the tubes are made from the transparent
material one can see the bubbles coming from membrane. Since the
pressure, when the bubbles start to come out, is far below from
the bubble point pressure (pressure which opens the largest pores
of the membrane) one concludes that bubble formation is due to the
gas diffusion flow.

There are at least two possibilities to correct the effect of gas
diffusion in liquid porosimetry tests. The first possibility is to
run the test as fast as possible when the test-time is less then
the time for significant effect of gas diffusion. But that is
possible under very special condition such as high liquid flow
rate through the system sample-membrane. The second possibility is
to study the bubble formation effect of the given membrane and
then by knowing that information try to subtract the bubble
formation from the final data.  Practically that means to run just
the membrane as a blank test then to subtract the blank data from
the final sample data. Both these possibilities are useful but
they cannot solve the problem radically.

\subsection*{C. Description of the invention}

\begin{figure}[htb]
\vspace*{1cm} \hspace*{1.5cm} \centerline{ \vbox{ \epsfysize=75mm
\epsfbox{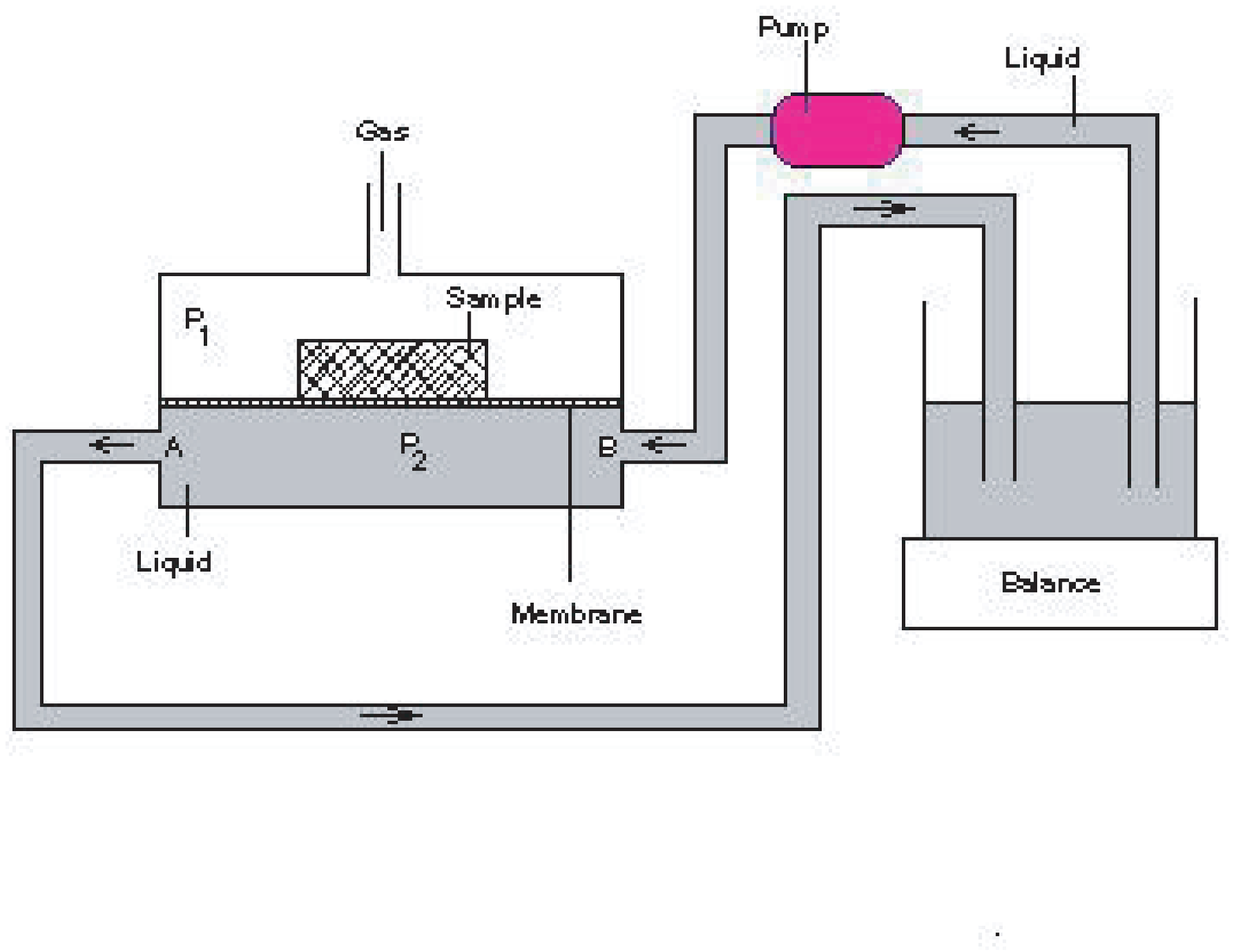} } } \vspace*{-0.5cm}\caption{Modified scheme of
liquid porosimetry. } \label{fig:3}
\end{figure}

To solve the problem of the gas diffusion bubble formation I
suggest a method of controlling gas diffusion and bubble formation
in liquid porosimetry. The essence of the method is the creation
of the liquid-flow bellow the membrane to wash out the gas bubbles
during the test. On Fig. 3 the scheme of the method is shown. The
differences with standard liquid porosimetry presented on Fig.1
are obvious. I used the pump with variable flow. The pump can be
switch on in reverse mode. The direction of the flow is not
important here. During the test the pump was on constantly. The
bubbles wash out into the glass, which is placed on the
top-loading balance.  When pump is on it causes differential
pressure between points A and B below the membrane no more than
$0.05 -0.2$ PSI depending on flow through the pump. That causes an
error in real pressure reading of DP. To reduce this error I
recommend switching the pump to high pressure. For example at
$DP=10$PSI and $PA-PB=0.05$ PSI the error in pressure reading is
just $0.5\%$,
 at $DP=100$PSI and $PA-PB=0.2$PSI the error is $0.2\%$.
 At low pressure there is no need to have a liquid flow
 bellow membrane because the rate of bubble formation is not significant.

\begin{figure}[htb]
\vspace*{-1.0cm} \hspace*{1.5cm} \centerline{ \vbox{
\epsfysize=110mm \epsfbox{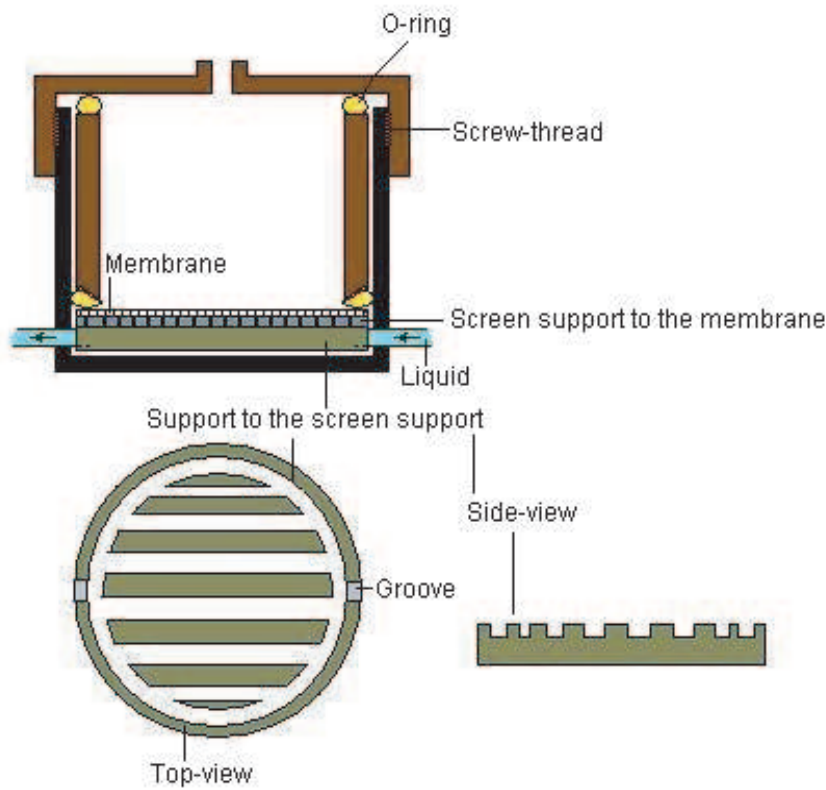} } }
\vspace*{-1cm}\caption{Scheme of sample chamber. } \label{fig:4}
\end{figure}

The {\bf actual realization} of the method is the following: The
membrane is placed on thin metal screen support (I used Millipore
$47$mm SST Support Screen). To prevent the deformation of the
screen under the pressure another support is placed under the
first screen-support. The last one contains parallel grooves to
direct the liquid flow (see Fig. 4) and two holes for incoming and
out-coming tubes. The pump used was Fisher brand Variable-Flow
Peristaltic Pump with ability to regulate the liquid flow from
$4.0$ up to $85.0$ mL/min. All other features are the same in
principle as in standard liquid porosimetry \cite{MT,RM}.

\section{Conclusion}

The suggested method allows controlling the diffusion through
membrane. It allows the using membrane up to bubble point pressure
and even more, if the gas flow through membrane just after the
bubble point is very small (several cc/min.). For example, with a
milli-pore $0.025$ micron (mm) membrane using Galwick as a liquid
it is possible to see $0.1$ micron pore sizes in diameter, which
correspond to up $100$-PSI pressure.

Using the considered scheme and method the liquid porosimetry
overcomes its current physical limits pointing to the left problem
of finding better membranes.



\noindent
\begin{thebibliography}{99}
\bibitem{MT}
B. Miller, I. Tyomkin, Journal of colloidal and interface science
162, 163-170 (1994).
\bibitem{RM}
L. Rebenfeld, B. Miller, J.Text.Inst., 1995, 86 No.2@Textile
Institite.
\end{thebibliography}
\end{document}